# Temperature and Rate Dependent Constitutive Behaviors of Low Melt Field's Metal

Quang-Kha Nguyen, Fanghang Deng, Pu Zhang[*]

*Department of Mechanical Engineering, State University of New York at Binghamton, Binghamton, NY 13902*

**Abstract**

Low melting point metals such as Field's metal and gallium are increasingly used as transition phases in stiffness-tuning soft materials and devices. Nevertheless, there is a knowledge gap on the fundamental constitutive behaviors of metals with melting points below 100°C. This letter reports the stress-strain relationships of Field's metal at various temperatures and strain rates. This metal has the lowest melting point among metals whose constitutive behaviors have been studied systematically. Experimental results indicate that Field's metal exhibits a strain-softening behavior, in oppose to the strain-hardening behaviors of most engineering metals and alloys. A modified Johnson-Cook model is devised in this letter to describe the constitutive behaviors of Field's metal. The proposed model will facilitate the design and analysis of functional materials and structures consisting of Field's metal.

## 1. Introduction

Low melting point metals (LMPM) [1] have been employed to fabricate functional materials and structures [2–11] with tunable stiffness, shape memory effect, self-healing behaviors, good thermal/electrical conductivity, etc. The most commonly used lead-free LMPMs include gallium (melting point $T_m$ = 30°C) and Field's metal ($T_m$ = 62°C), which are also called liquid metals sometimes [1]. The most prominent feature of LMPMs is their solid-liquid phase transition at relatively low temperatures, e.g. gallium melts at normal body temperature. This phase transition is accompanied with a loss of stiffness and the fusion of defects due to the melting process. Such characteristics can be employed to design and fabricate LMPM composites with tunable stiffness and self-healing behaviors [2–4,7,9]. Besides, a shape memory effect is attained if the LMPM composites are hot/cold programmed and re-melted [2,3,5,6,9,10], similar to the behavior of shape memory polymers. These fascinating characteristics have given rise to a big class of multifunctional materials based on LMPMs.

Although LMPMs have not drawn much attention until recently, early works can be traced back to early 2000s. Nakai et al. [12] fabricated a metamorphic robot with tunable rigidity and shape by embedding a lead-based LMPM ($T_m$ = 47°C) into a silicone rubber. Whitesides et al. [8] proposed an idea of microsolidics and fabricated LMPM mesh structures embedded in a silicone matrix. Later on, a few researchers fabricated stiffness-tuning devices and actuators by using either lead-based LMPM [13] or Field's metal [4,7,14]. Shepherd et al. [2] gave the first comprehensive introduction to the stiffness tuning, shape memory, and self-healing behaviors of LMPM composites. This work has led to a surge of multifunctional materials based on LMPMs. For example, composites comprising of Field's metal particles and polymers have been synthesized and tested for their shape memory effect [5,9], stiffness-tuning ability [3], toughness enhancement [15], and self-healing behavior [9]. Moreover, researchers also studied the mechanical characteristics of gallium-based composites [6,10,16], which are similar to that of Field's metal composites. Recently, the authors have developed multifunctional lattice materials [11] based on Field's metal, which exhibit intriguing properties such as recoverable energy absorption, tunable stiffness, and reconfigurable shapes.

---

[*] Corresponding author. Email address: pzhang@binghamton.edu

There is a pressing need to investigate the constitutive behaviors of LMPMs for at least two reasons. (1) LMPMs are increasingly used in functional materials and structures either in the form of particles or structured components. Design of these materials and structures requires substantial efforts to characterize and model the constitutive behaviors of LMPMs. However, there is a lack of research on experimental testing and modeling to fulfill this need. For example, the current knowledge of Field's metal is merely on the elastic modulus [4] and yield strength [17], which is insufficient for advanced engineering analysis and design. (2) The constitutive behavior of LMPMs remains to be a knowledge gap. Researchers have studied the stress-strain relationships of engineering metals and alloys with high melting points [18] and soldering alloys with intermediate melting points [19]. Nevertheless, there is little understanding of the constitutive behaviors of metals and alloys with melting points below 100°C. Therefore, exploring the constitutive behaviors of LMPMs helps fill this knowledge gap in the range of extremely low melting points.

The purpose of this letter is to characterize and model the constitutive behaviors of LMPMs at different temperatures and strain rates. Taking Field's metal as an example, we conduct uniaxial tensile and compression testing on as-casted specimens. The stress-strain relationships are described by devising a modified Johnson-Cook model. The model is verified by comparing the experimental and simulation results of a honeycomb lattice structure subject to uniaxial tension. The proposed model can be used to simulate functional composites and structures consisting of Field's metal.

## 2. Experimental Results

The experimental procedure of the uniaxial tensile and compression testing is introduced first. All testing specimens were prepared by casting molten Field's metal (Rotometal Inc.) into silicone molds. The gauge section of the tensile specimen is 4.25 mm × 25 mm with a thickness of 1.5 mm; while the compression cylinders are 10 mm in diameter and 10 mm tall. The uniaxial testing is performed on a load frame (MTS 858, 5000 N load cell) equipped with an environmental chamber (MTS 651). An extensometer (Epsilon, 0.5 inches gauge) is installed on the specimens during the tensile testing to measure the strain accurately. For compression testing, Teflon tapes are placed in between the Field's metal cylinder and compression platens to reduce friction. In order to characterize the temperature and rate dependent constitutive behaviors of Field's metal, we have tested the tensile and compression specimens at five temperatures ($T = 23°C, 35°C, 45°C, 50°C, 55°C$) and three strain rates ($\dot{\varepsilon} = 0.002$ s$^{-1}$, 0.01 s$^{-1}$, 0.05 s$^{-1}$). At least three samples were tested for each set of testing parameters.

The constitutive behaviors of Field's metal are introduced and discussed below. The stress-strain curves obtained from our uniaxial testing are presented in Fig. 1 for compression and Fig. S1 for tension (see Supplementary Information). Note that the measured stress-strain curves in the tensile testing are only reliable for small strains because necking occurs during the post-yielding process. Therefore, the compression testing results are more reliable for the post-yielding responses. It is observed in Fig. 1 that the Field's metal exhibits a strain-softening behavior after yielding with the flow stress decays to a steady-state plateau stress. To help understand these behaviors, we also compared the elastic modulus, yield stress, and plateau stress values in Fig. 2. The following characteristics are observed for the constitutive behaviors of Field's metal:

1) *Strain-softening*. It is observed from Fig. 1 that Field's metal does not have any strain-hardening effect, which is uncommon for engineering metals and alloys. The flow stress in the post-yielding stage decreases gradually until it reaches a plateau stress, which implies that the plastic deformation can alter the microstructure and reduce the flow resistance. Such a strain-softening behavior was also found for gallium ($T_m$ = 30°C) [6], Wood's metal ($T_m$ = 71°C) [20], and Lichtenberg's metal ($T_m$ = 92°C) [20]. Thus, we anticipate that strain-softening is a very common feature of LMPMs with melting points below 100°C.

2) *Pressure-insensitive*. The yield stress values obtained from the tensile and compression testing are compared in Fig. 2 (b). It is observed that the tensile yield stress is close to that measured in the compression testing, although the latter is usually slightly larger. We conclude from this comparison that the yielding behavior of Field's metal is pressure-insensitive, similar to most other metals.

3) *Soft, fragile, yet ductile*. LMPMs are usually softer and more fragile than engineering metals by considering their relatively low stiffness and strength. Some researchers anticipate that they are very brittle, too. However, we found from the tensile testing that Field's metal can be drawn to 30% of true strain without failure (see Fig. S2). Thus, Field's metal is soft, fragile, yet ductile. Future research is needed to enhance the stiffness and strength of Field's metal, for instance, by adding nano-fillers.

4) *Temperature and rate dependent*. Based on the data in Fig. 1 and Fig. 2, Fields' metal exhibits a viscoelastic-viscoplastic behavior that is strongly dependent on temperature and strain rates. The elastic modulus, yield stress, and plateau stress all change dramatically once the temperature and strain rate are changed. This phenomenon is not surprising by considering that the testing temperature is close to its melting point. As shown in Fig. 2 (a), the elastic modulus of Field's metal can be fitted into a quadratic function in terms of the strain rate and temperature, as

$$E(\dot{\varepsilon}, T) = 1.059 \times 10^4 + 1.055 \times 10^5 \dot{\varepsilon} - 8.443T - 1.107 \times 10^6 \dot{\varepsilon}^2 + 301.6\dot{\varepsilon}T - 2.543T^2 \quad (1)$$

where the units of the modulus $E$, strain rate $\dot{\varepsilon}$, and temperature $T$ are MPa, s$^{-1}$, and °C, respectively. Equation (1) can be used at the temperature range of $20°C < T < 62°C$. For temperatures out of this range, more testing is needed to verify its accuracy. The yield stress and plateau stress can be fitted in similar formulae.

5) *Coupled strain-rate-temperature effect*. Most viscoplastic models assume that the effects of strain, strain rate, and temperature on the flow stress can be decoupled. Mathematically, this assumption is expressed as $\sigma = f_1(\varepsilon) f_2(\dot{\varepsilon}) f_3(T)$, where $\sigma$ is the flow stress, $f_1$, $f_2$, and $f_3$ represent the influences of strain, strain rate, and temperature, respectively [21]. For example, the renowned John-Cook model [22] and Anand model [23] follow this approach. This decoupled model actually rescales the whole stress-strain curve by a constant factor once the strain rate or temperature is changed. However, this assumption is not true for Field's metal. Based on the observation from Fig. 1, Fig. 2 (b), and Fig. 2 (c), changing either the strain rate or temperature will rescale the flow stress, but the rescaling factor depends on the strain level. Therefore, we conclude that the strain-rate-temperature effects cannot be decoupled for Field's metal.

6) *Constant initial-softening-slope*. As shown in Fig. 1, the softening slope in the stress-strain curves decreases from an initial value to zero when the strain increases. In the literature, the initial-softening-slope usually declines when the strain rate decreases or the temperature increases, e.g. for glassy polymers [24,25] and metallic glass [26]. By evaluating the initial-softening-slopes of Field's metal (see Fig. S3), we found that the initial-softening-slope is almost independent of the strain rate and temperature. The physical origin of this abnormal behavior is still unclear.

These complicated features make it rather difficult to establish an appropriate constitutive model for Field's metal, which will be addressed in Section 3.

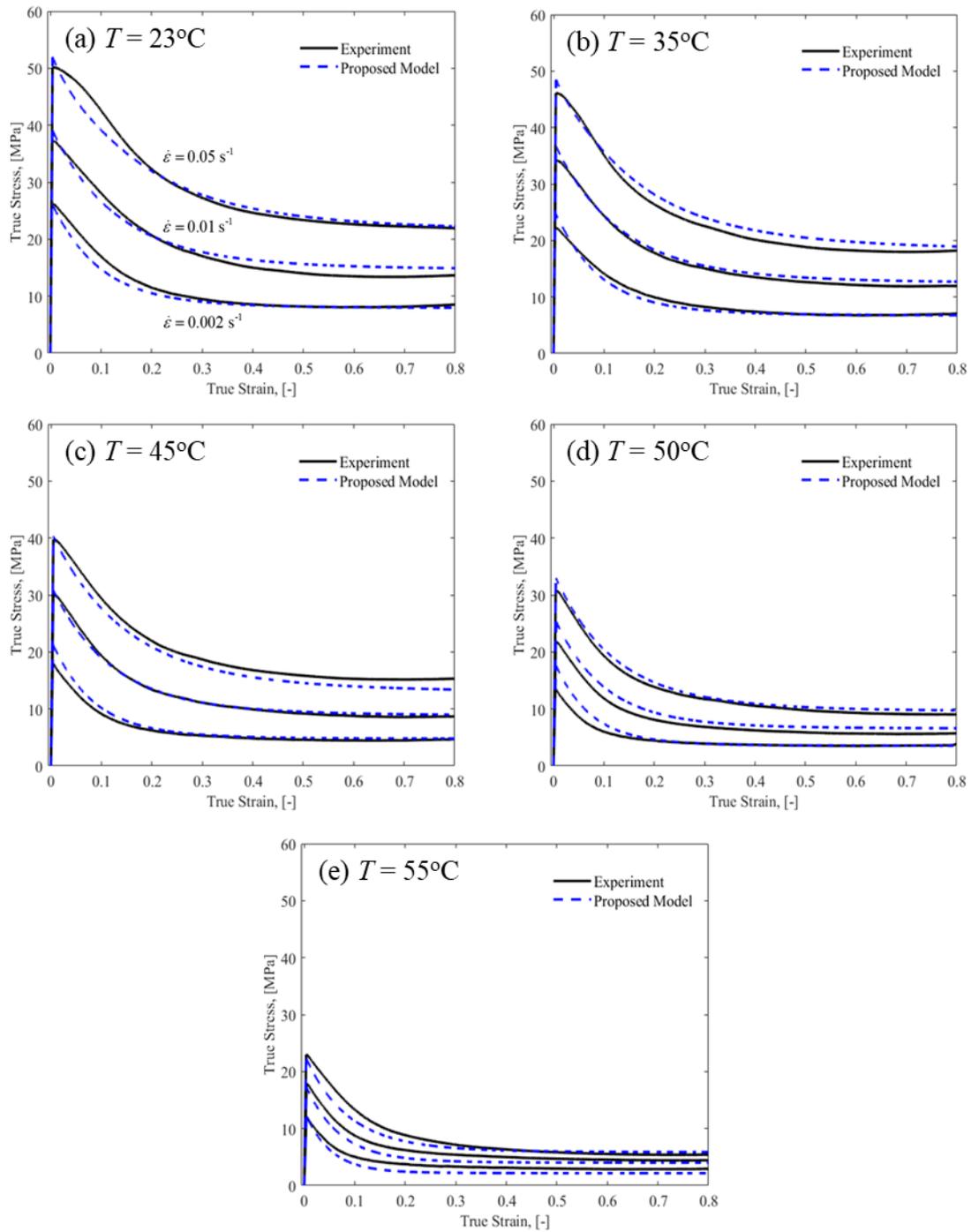

Fig. 1. Stress-strain curves of Field's metal under uniaxial compression testing. Both experimental and simulation results are illustrated.

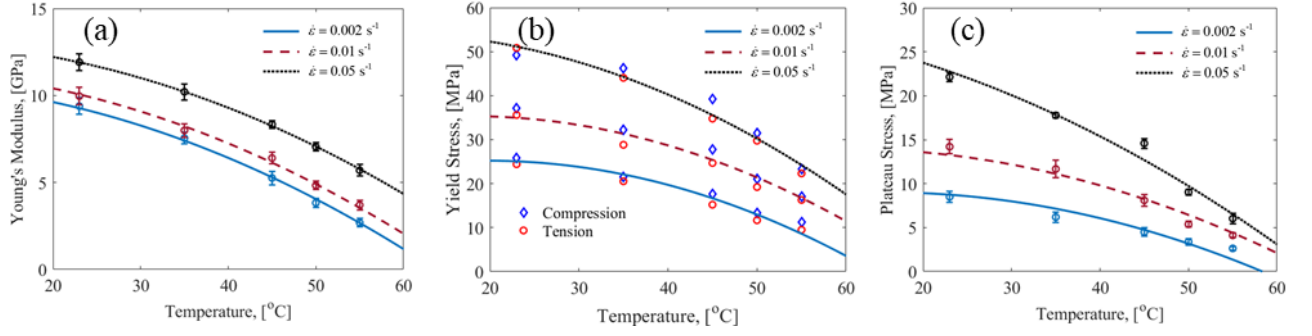

Fig. 2. Temperature and rate dependent mechanical properties of Field's metal: (a) Elastic modulus; (b) Yield Stress; (c) Plateau stress.

## 3. Proposed Model

We propose a new constitutive model to describe the stress-strain relationships of Field's metal. The Johnson-Cook model [22], Anand model [23], and Arrhenius model [27] are among the most widely used constitutive models for metals that depend on temperature and strain rate. The Johnson-Cook model and Anand model are convenient to use but not suitable for Field's metal because they do not consider the coupled strain-rate-temperature effect and constant initial-softening-slope. In contrast, the Arrhenius model is able to take into account these effects, but suffers from its complexity and large number of material parameters. Therefore, we devise a facile model for Field's metal by modifying the Johnson-Cook model.

In order to model the strain-softening and coupled strain-rate-temperature effect, we decompose the flow stress $\sigma$ into a transient flow resistance $S$ and a steady state plateau stress $\sigma_s$, as

$$\sigma = S + \sigma_s \tag{2}$$

A graphical illustration of this stress decomposition is shown in Fig. 3. The transient flow resistance $S$ characterizes a diminishing stress barrier that resists the plastic flow. This stress decomposition was adopted by Anand et al. [24,26] for both polymers and metals. The transient flow resistance $S$ and plateau stress $\sigma_s$ are modeled independently to account for the strain-softening and coupled strain-rate-temperature effect.

We assume that the plateau stress $\sigma_s$ follows the Johnson-Cook model, as

$$\sigma_s = A\left(1 + C_1 \ln \frac{\dot{\varepsilon}}{\dot{\varepsilon}_o}\right)\left[1 - \left(\frac{T - T_r}{T_m - T_r}\right)^{m_1}\right] \tag{3}$$

where $\dot{\varepsilon}_o$ is a reference strain rate, $T_m$ is the melting point, and $T_r$ is a reference temperature. There are three material constants to fit, i.e. $A$, $C_1$ and $m_1$. Similar to the Johnson-Cook model, the $C_1$ and $m_1$ terms characterize the rate and temperature effect, respectively. Note that the strain-hardening term in the original Johnson-Cook model is ignored.

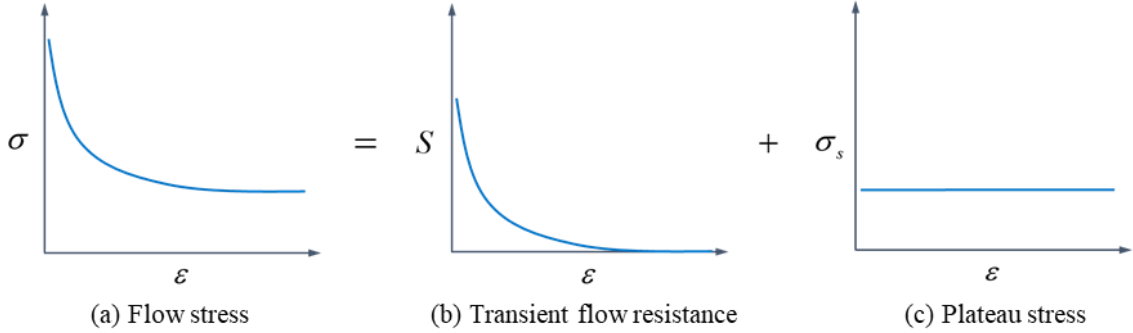

(a) Flow stress     (b) Transient flow resistance     (c) Plateau stress

Fig. 3. Schematic illustration of the flow stress decomposition for Field's metal.

The transient flow resistance $S$ is modeled in a different way by considering that the initial-softening-slope $dS/d\varepsilon$ is constant at $t=0$. We use a differential equation to predict the evolution of $S$, as

$$\dot{S} = -h_o \dot{\varepsilon} \frac{S}{S_o} \tag{4}$$

where $h_o$ is the initial-softening-slope, and $S_o = S(0)$ is the initial value of the transient flow resistance $S(t)$. The initial resistance $S_o$ also follows the Johnson-Cook model, as

$$S_o = \bar{S}_o \left(1 + C_2 \ln \frac{\dot{\varepsilon}}{\dot{\varepsilon}_o}\right) \left[1 - \left(\frac{T-T_r}{T_m-T_r}\right)^{m_2}\right] \tag{5}$$

where $\bar{S}_o$, $C_2$, and $m_2$ are three materials constants to be determined. The constant initial-softening-slope is incorporated in Eq. (4) automatically. At $t=0$, Eq. (4) degenerates to $\dot{S} = -h_o \dot{\varepsilon}$ by setting $S = S_o$, which yields a constant initial-softening-slope as $dS/d\varepsilon = -h_o$. This differential equation technique has been used to model the strain-softening of polymers and metals in the literature [24–26]. However, their evolution equation does not yield a constant initial-softening-slope.

## 4. Calibration and Verification

There are totally seven plasticity constants to calibrate in Eqs. (3)-(5). The two parameters $A$ and $\bar{S}_o$ can be determined readily by fitting the plateau stress and initial transient flow resistance at the reference temperature $T_r = 23°C$ and reference strain rate $\dot{\varepsilon}_o = 0.002 \text{ s}^{-1}$. The remaining five constants $C_1$, $m_1$, $h_o$, $C_2$, and $m_2$ are determined by minimizing the difference between the stress-strain curves obtained from experiment and the proposed model. For Field's metal, the calibrated parameters are shown in Table 1. Figure 1 compares the stress-strain curves of Field's metal obtained from both experiment and the proposed model. It is found that the proposed model fits the experimental data very well. The model accounts for the strain-softening, coupled strain-rate-temperature effect, and constant initial-softening-slope. In addition, with only seven plasticity parameters involved, this model is much more convenient to use compared to the Arrhenius model. We expect that this new model is applicable to many LMPMs other than Field's metal.

Table 1. The calibrated plasticity parameters of the proposed model for Field's metal.

| $A$ (MPa) | $\bar{S}_o$ (MPa) | $C_1$ | $m_1$ | $h_o$ (MPa) | $C_2$ | $m_2$ |
|---|---|---|---|---|---|---|
| 7.9204 | 18.3291 | 0.5267 | 1.6152 | 205.4702 | 0.2077 | 3.9377 |

In order to further verify the reliability of the proposed model, we conducted uniaxial tensile testing on honeycomb lattice structures made of Field's metal and compared with the results from simulation. We fabricated honeycomb structures by casting molten Field's metal into a 3D printed polymer mold (Flashforge printer). The struts in the honeycomb structure have rectangular cross-sections that are 40 mm long, 20 mm wide, and 1.5 mm thick. Specimens are tested at $T = 23°C$ and $T = 50°C$ with two strain rates $\dot{\varepsilon} = 0.01 \text{ s}^{-1}$ and $\dot{\varepsilon} = 0.05 \text{ s}^{-1}$. The obtained force-displacement curves are illustrated in Fig. 4 (b). It is observed that the whole structure shows a strain-softening behavior, similar to bulk Field's metal. The simulation is carried out in ABAQUS 2019 [28] after implementation of the constitutive model. The structure is discretized by hexahedral elements (C3D8R) with mesh size that meets the convergence criterion. The Poisson's ratio of Field's metal is set as 0.33. The simulated force-displacement curves are compared to the experimental data in Fig. 4 (b). Overall, the proposed constitutive model can predict the large inelastic deformation and softening of the honeycomb structures very well at different strain rates and temperatures. As shown in Fig. 4 (b), we have observed some discrepancy in the linear regimes of the force-displacement curves given the fact that the experimental stiffness is smaller than the simulation one. This discrepancy is attributed to testing error since the initial deformation stage involves stiffening of the fixtures.

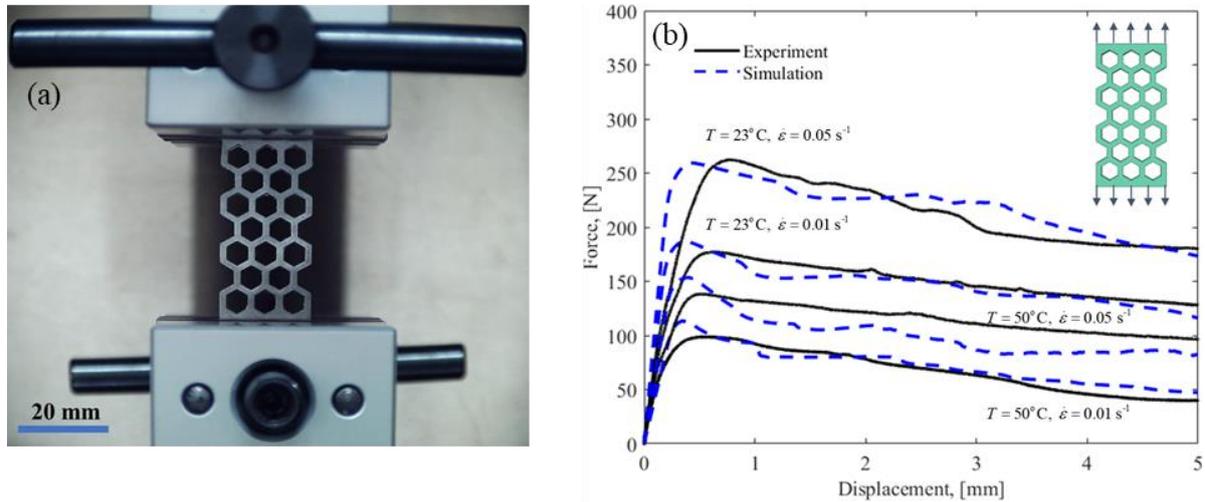

Fig. 4. Tensile testing of honeycomb lattice structures made of Field's metal: (a) The experimental set up; (b) Force-displacement curves of the honeycomb structures.

## 5. Conclusions

Low melting point metals (LMPM) have gained more and more applications in functional materials and structures. However, there is scarce knowledge on the fundamental constitutive behaviors of LMPMs, which is an obstacle for advanced design and analysis. This letter addresses this issue by investigating the stress-strain relationships of Field's metal, one of the most commonly seen LMPMs, and devising a new constitutive model to describe its constitutive behavior. With a typical elastic modulus around 10 GPa, Field's metal is softer than most engineering metals but much stiffer than engineering plastics. Experimental results indicate that the Field's metal exhibits a strain-softening behavior. In particular, the initial-softening-slope is independent of temperature and strain rate, which is different from most metals and polymers studied in the literature for decades. The stress-strain relationships are strongly dependent on the strain rate and temperature mainly because the melting point is close to room temperature. In addition, the effects of strain, strain rate, and temperature on the flow stress cannot be decoupled like most viscoplastic models. In order to describe all these complicated constitutive behaviors, we have developed a new constitutive model for Field's metal. The accuracy and reliability of the proposed model are verified by comparing the experimental and simulation results on compression cylinders and honeycomb lattice structures made of Field's metal.

In brief, the current work fills a knowledge gap on the constitutive behaviors of metals with melting points below 100°C. In addition, the proposed constitutive model will be useful for future simulation and design regarding LMPMs. We look forward to seeing more research on the constitutive behaviors of other LMPMs (e.g., gallium) and simulation studies on LMPM materials and structures.

## 6. Acknowledgements

This work is supported by the start-up fund from the Waston School of Engineering and Applied Science at SUNY Binghamton.